# Semantic Gateway as a Service architecture for IoT Interoperability

Pratikkumar Desai[1], Amit Sheth, Pramod Anantharam

Ohio Center of Excellence in Knowledge-enabled Computing ([Kno.e.sis](Kno.e.sis)), Wright State University

## Abstract

The Internet of Things (IoT) is set to occupy a substantial component of future Internet. The IoT connects sensors and devices that record physical observations to applications and services of the Internet[1]. As a successor to technologies such as RFID and Wireless Sensor Networks (WSN), the IoT has stumbled into vertical silos of proprietary systems, providing little or no interoperability with similar systems. As the IoT represents future state of the Internet, an intelligent and scalable architecture is required to provide connectivity between these silos, enabling discovery of physical sensors and interpretation of messages between the things. This paper proposes a gateway and Semantic Web enabled IoT architecture to provide interoperability between systems, which utilizes established communication and data standards. The Semantic Gateway as Service (SGS) allows translation between messaging protocols such as XMPP, CoAP and MQTT via a multi-protocol proxy architecture. Utilization of broadly accepted specifications such as W3C's Semantic Sensor Network (SSN) ontology for semantic annotations of sensor data provide semantic interoperability between messages and support semantic reasoning to obtain higher-level actionable knowledge from low-level sensor data.

*Note to the reviewers: Unlike traditional academic journal publications, IEEE IC has a preference to limit number of references, so we welcome any suggestions on removing references, especially if additional references are suggested for inclusions. We have also included some introductions to communication technologies and an overview on current IoT ecosystem to make this manuscript as self contained as possible, especially for IC's wider audience. However, if needed, some of these can be removed if we want to assume that readers will be; actionable suggestions and recommendations on these matters will be valuable.*

## 1. IoT Interoperability crisis

In the initial momentum of IoT, smart grid, smart appliances, and wearable device powered health and fitness are emerging as major application domains but with varying architecture and data models. Figure 1 shows vertical silos for these domains with examples including physical sensors to the Internet service. In health care domain, the Fitbit, an activity-monitoring device, provides complete sets of IoT components creating its close silo. It provides graphical interface and uses representational state transfer (REST) application interface to connect the sensor to their cloud service. Similarly, a user can connect and monitor his health by analyzing data from sensors such as heart rate, glucose, weighing scale using any popular open hardware platform such as Raspberry Pi or Arduino as a gateway node. An IoT service such as Xively, previously known as Pachube, can provide graphical interface for sensor data aggregated from this gateway node. The current state of IoT infrastructure lacks methods to provide interconnectivity, for example between the Fitbit and the Xively silos, at each of these layers: Network, Messaging and Data model.

---

[1] Pratikkumar Desai is now with SeerLabs, an IoT startup.

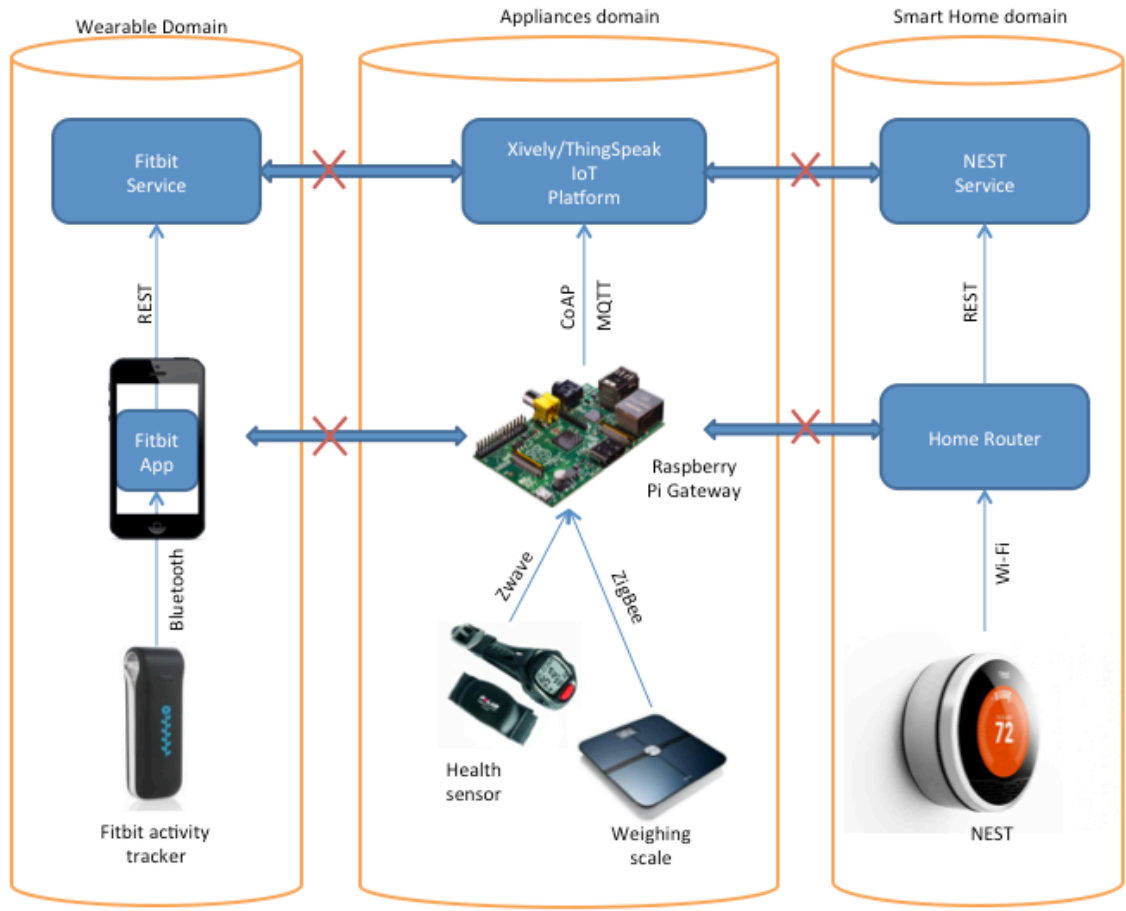

*Figure 1: Vertical silos of IoT service deployment*

## 1.1 Network layer interoperability

The power constrained sink nodes, connected to the physical world objects, require efficient networking protocols. The IoT domain is scattered between various low power networking protocols (ZigBee, ZWave, and Bluetooth), traditional networking protocols (Ethernet, WiFi) and even hardwired connections. Figure 2 shows IoT networking protocols with traditional devices associated with them. These protocols are designed for domain specific applications with distinctive features. Solving interoperability issue at this level requires standardization at the hardware level. Various commercial products have been developed to support multiple networking protocols by assembling the required hardware components together. This paper reports on the research on solving the interoperability problem at the application level, bypassing the networking protocol interoperability challenge.

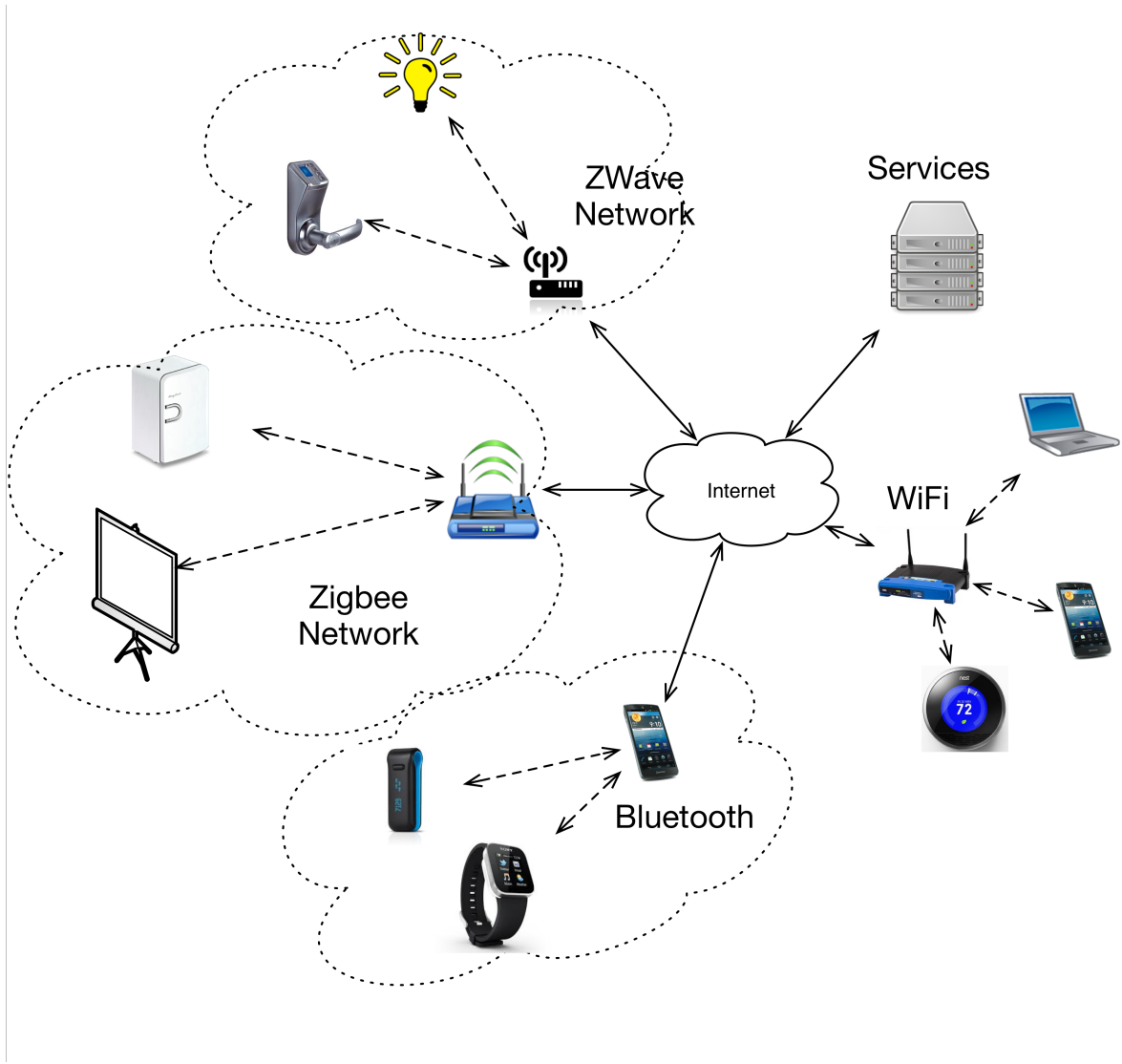

*Figure 2: Present state of IoT network architecture*

## 1.2 Interoperability between messaging protocol

In contemporary IoT applications, multiple competing application level protocols such as CoAP (Constrained Application Protocol), MQTT (Message Queue Telemetry Transport) and XMPP (Extensible Messaging and Presence Protocol) are proposed by various organization to become the de facto standard to provide communication interoperability[2]–[4]. Each of the protocol possesses unique characteristics and messaging architecture helpful for different types of IoT applications, which require effective utilization of limited processing power and energy. However, a scalable IoT architecture should be independent of messaging protocol standards, while also providing integration and translation between various popular messaging protocols.

Figure 3 shows an example of REST based messages transfer between CoAP client and server, while the Figure 4 shows publisher/subscriber based message delivery for MQTT protocol.

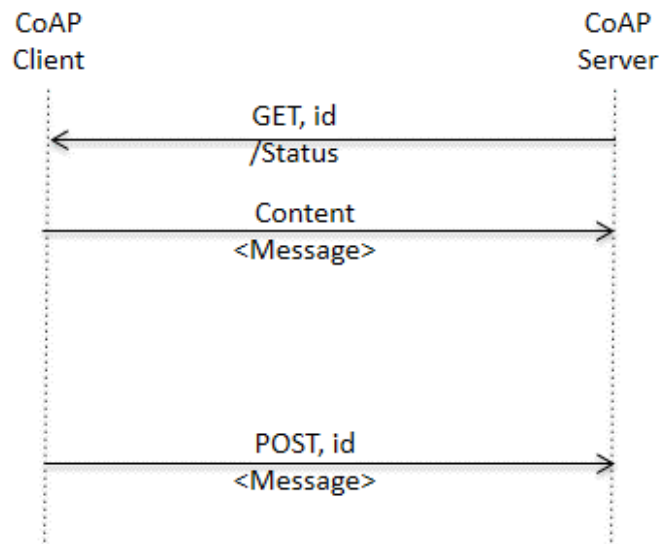

*Figure 3: An example of CoAP message transfer*

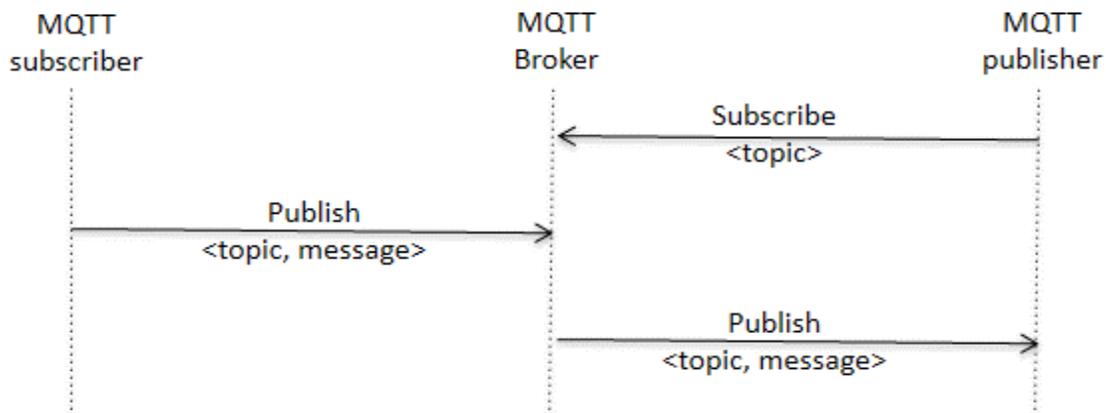

*Figure 4: An example of publisher-subscriber based message transfer*

We describe a semantic IoT architecture where the gateway, located between physical level sensors and cloud-based services, provides translation between widely used CoAP, MQTT and XMPP protocols, making their semantic integration possible and seamless.

## 1.3 Interoperability at data annotation level

The traditional paradigm of the IoT service model provides raw sensor data to the software agent, captured from the heterogeneous sink nodes. This raw sensor data do not contain any semantic annotation and requires extensive manual effort in order to build practical applications.

An IoT service can provide raw sensor data with added metadata but due to absence of annotation standards, it cannot be exploited by other services. Typically IoT applications are deployed in a bottom-up (sensors, gateways, service and application) manner from a common provider. These providers control the sensor data and data structures, which help them to create intelligent application on top of it. Due to the proprietary approach employed by these providers, the IoT domain has turned into a domain of vertical silos of various IoT applications with no horizontal connectivity between them. This lack of interoperability with independent services currently endangers the wide acceptability and adoption of the IoT domain, especially for applications that can benefit from multiple devices.

## 2. Background

IoT interconnects physical world "Things" by utilizing software and networking technologies. Due to its roots in traditional sensor networks, connected physical objects are resource-constrained devices, and require competent communication protocol for energy efficiency.

First wave of IoT application in smart city domain emphasized on connecting sensor interfacing with physical-world using lightweight protocols such as CoAP and XMPP [5][6]. In later stages, traditional Internet state transfer protocol such as REST is used for similar applications, where event-centric frameworks had been implemented to reduce number of messages transmitted [7]. The 'Smart-Object' devices with domain specific intelligence are rapidly replacing first wave of IoT devices [8]. Although these devices do not utilize semantic technologies, they provide higher-level of awareness from the sensor than just plain raw sensor data.

The IoT domain has been started getting congested with heterogeneous applications using different communication protocols and data models [9]. Various organizations such as the OpenIoT alliance, AllSeen alliance, and IPSO alliance are working on standardization of communication protocols to provide interoperability between various vendors silos [10][11][12]. Organization such as Internet Engineering Task Force (IETF) and XMPP standards foundation are trying to scale their messaging protocols, CoAP and XMPP, respectively, to align with other protocols. These efforts are scattered and largely focus on solving problems around one protocol instead of providing integration solution.

In Web-centric infrastructure, acquisition of contextual information from raw sensor data requires annotation of sensor data with semantic metadata. Key standardization efforts that have sought to establish sensor data models for sensors to be accessible and controllable via the Web include:

- OGC Sensor Web Enablement (SWE)
  The SWE efforts established by the Open Geospatial Consortium include following important specifications: Observation & Measurement (O&M), Sensor Model Language (SensorML) and Sensor Observation Service (SOS)[13]. The O&M and SensorML contain standard model and XML schema for observations/measurements and sensors/processes respectively. The SOS is a standard service model, which provides mechanism for querying observation and sensor metadata.

- Semantic Sensor Network (SSN) ontology

    The SSN ontology, developed by W3C provides a standard for modeling sensor devices, sensor platforms, knowledge of the environment and observations[14] [15]. The SSN provides a foundation in the direction of achieving interoperability between the interconnected IoT Silos.

- Semantic Sensor Observation Service (SemSOS)

    The Semantic Web enabled implementation of SOS, SemSOS, provides a rich semantic backend (knowledge base) while retaining the standard SOS specifications/service interactions. A semantically intelligent client can utilize this capability of SemSOS to derive higher level abstractions from the annotated sensor data [16] by implementing a semantic reasoning service acting on the knowledge base. SemSOS is the principal component of Semantic Sensor Web [17].

Although the utilization of these standards provide integration of Semantic Web with sensor applications, the interoperability challenges on IoT is far from being solved and a semantic IoT architecture is required to provide interoperability between connected IoT systems. This architecture should support multiple IoT protocols and severe resource and energy constrains. One of the major initiatives, which utilizing Semantic Web for IoT architecture, includes the OpenIoT project, funded by Europe Union's framework program. The OpenIoT focuses on developing open source middleware for IoT interoperability using linked sensor data[10].

In standard IoT applications the sink nodes are energy-constrained devices and utilizes minimum resources to conserve the energy. Various proposals seek to optimize the resources and provide translation between application layer protocol via the gateway devices[9][6]. These approaches fail in achieving interoperability at defining sensor annotation model, which is required to provide service level interoperability between IoT systems.

## 3. Semantic IoT Architecture

In the present IoT ecosystem, various IoT components can be broadly categorized into three classes: sink nodes, gateway nodes, and IoT services. Typical sink nodes consist of household appliances or sensors observing the physical environment, which possess low computational resources, stringent energy constraints and limited communication resources. The gateway node works as a sensor data aggregator and provides connectivity with other sink nodes and service providers. The gateway nodes have more computing resources compared to the sink nodes and occasionally provide replacement for the sink nodes. The IoT services collect data from the various gateway nodes and provide user or event specific services using a graphics interface, a notification or application.

Although they consist of each components mentioned above, the current IoT silos only provide end-to-end message delivery and lacks accessibility to semantic data. Organizations such as IETF, which manages CoAP standards, and XMPP are working on standardizing sensor data models as steps toward semantic data annotation[18]. In process of solving the data model interoperability problem in IoT silos, these efforts are advancing in direction of creating silos around these protocols, where these data models are protocol centric and incompatible with other data models.

Semantic annotation of sensor data by utilizing a standard mechanism and vocabulary can provide interoperability between IoT vertical silos. Semantic Web community has created and optimized standard ontologies for sensor observation, description, discovery and services via O&M, SensorML, SOS and SSN. By integrating these annotated data and providing Semantic Web enabled messaging interface, a third party service can convert heterogeneous sensor observations to higher level abstractions[19].

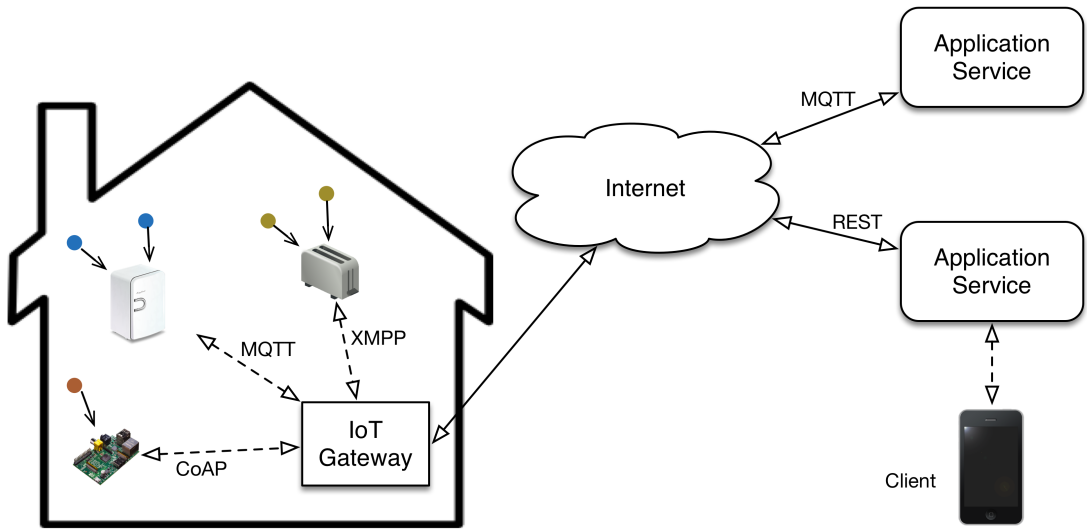

*Figure 5: Proposed IoT architecture with Semantic Gateway.*

Because gateway nodes have sufficient computational resources, we can implement necessary technologies to provide interoperability. Similarly, utilizing semantic technologies at the service level can also enable interconnection between them. We propose the concept of **Semantic Gateway as Service** (SGS) as a bridge between sink nodes and IoT services. In the proposed semantic IoT architecture, the gateway acts as the center of data communication between the physical-world and the Cloud. This architecture can be categorized as a Semantic Service Oriented Architecture (SSOA) for IoT systems as it fulfills technical requirements such as service-oriented architecture, standard based design, and semantic-based computing leveraging application agents to autonomously interpret sensor data and interact mutually [20][21].

The sink nodes can be connected to each other in a mesh or a hierarchical topology with wired or wireless connection. A node in the topology acts as the endpoint and connects to the gateway using CoAP, XMPP or MQTT protocol. Due to the lower processing capabilities of the sink nodes, they can be only utilized as clients. The CoAP protocol provides data in JSON or XML format while the MQTT only support XML. The data transferred from the sink nodes to the gateway is in raw format without any semantic annotations. As described in Figure 5, the SGS provides interfaces to Application services via REST and publisher/subscriber based protocols.

The data is semantically annotated at the gateway and hence these services can exploit the sensor information for further analysis.

This architecture is well suited to addresses privacy issues by allowing the users to control sensor data at the gateway, and hence may make it more acceptable. The gateway also implements high security standards by letting user specify the public and private sensor features, where private sensor features are only accessible after secure authorization using OAuth.

# 4. Semantic Gateway as Service (SGS)

The heart of the semantic IoT architecture is the SGS, which bridges low level raw sensor information with knowledge centric application services by facilitating interoperability at messaging protocol and data modeling level. The description below is complemented by Open Source code available at https://github.com/chheplo/node-sgs which is further being enhanced and evaluated in the context of CityPulse (http://www.ict-citypulse.eu/), a large multi-institutional EU FP7 supported project along with an effort for additional community engagement and development.

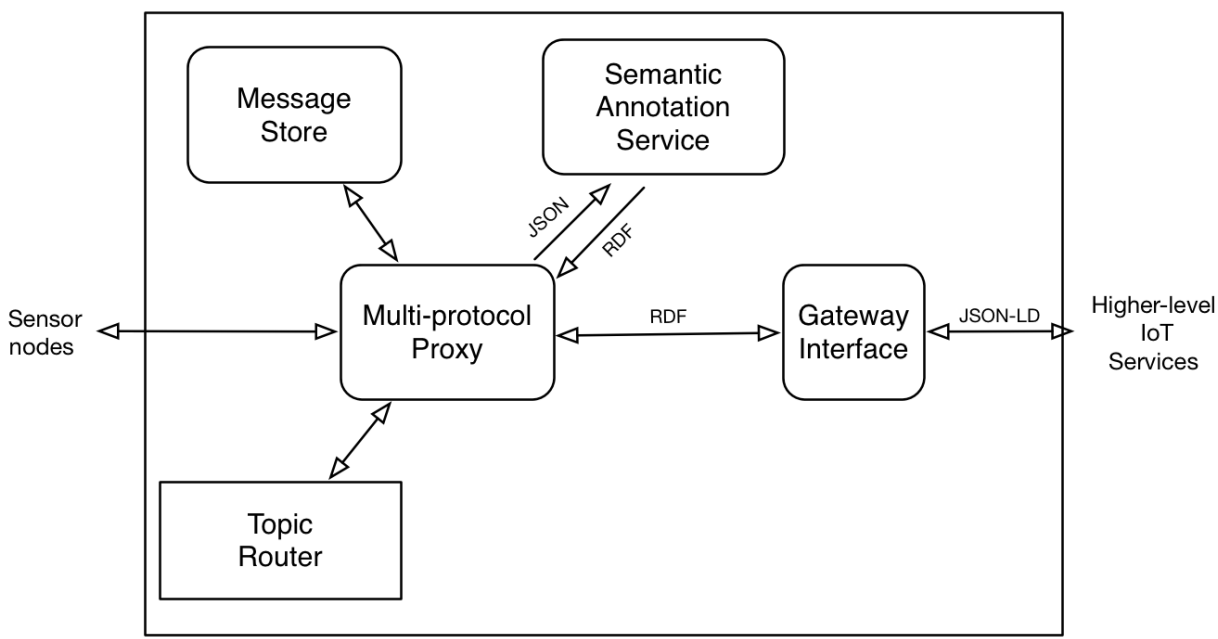

*Figure 6: SGS architecture.*

The SGS has three core components as described in Figure 6: (1) multi-protocol proxy, (2) semantic annotation service, and (3) gateway service interface. The SGS also has components for required capabilities such as message store and topics router, which assist multi-protocol proxy and gateway service interface in translation between messaging protocol. At a high level, SGS architecture connects external sink nodes to the gateway component using primitive client agents, which support MQTT, XMPP or CoAP. In contrast, the gateway service interface connects cloud services or other SGSs via REST or pubsub protocol. Before raw sensor data is forwarded from proxy to gateway interface, it is annotated using SSN and domain specific ontologies. Although the semantically annotated data is in RDF format at the multi-protocol proxy, the gateway interface converts the data into JSON, specifically linked data (JSON-LD) format to support RESTful protocols.

## 5. Multi-protocol proxy

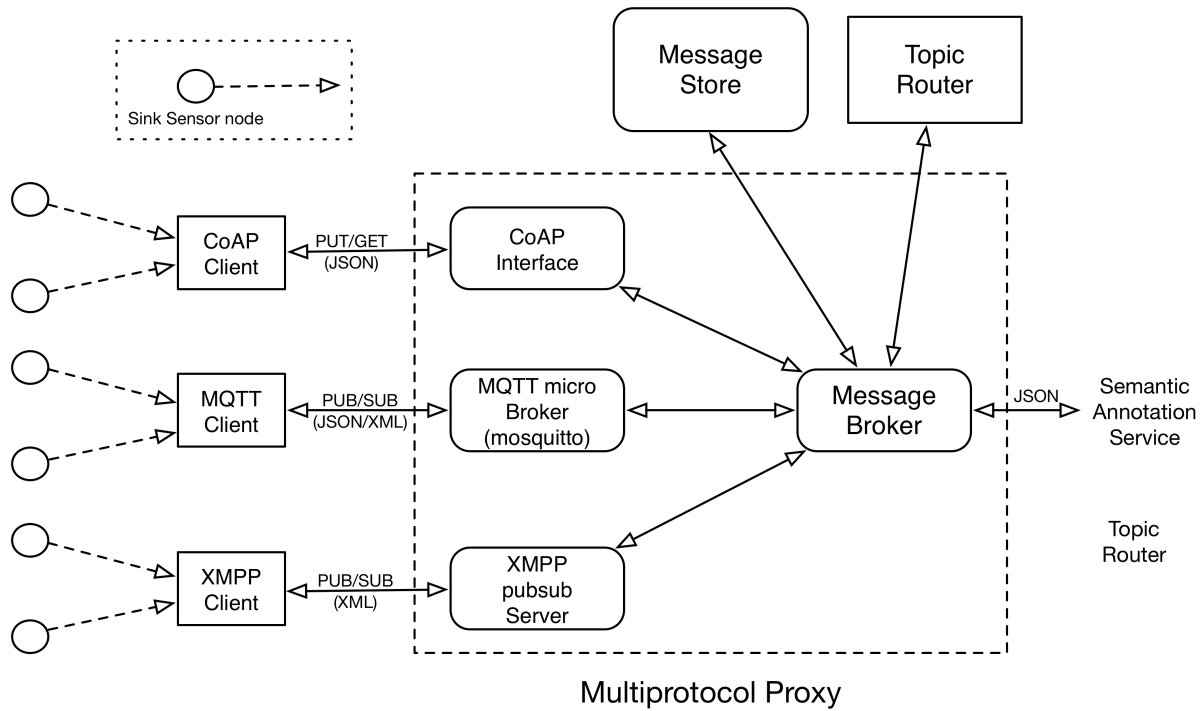

*Figure 7: Multi-protocol proxy, communicating with sensor nodes.*

The multi-protocol proxy is the SGS component facing the physical-world. Due to computation capability constrains, the sink level sensor nodes can support messaging protocols only as clients with limited support. CoAP is an optimized REST protocol for sensor applications, which supports request/response and resource/observer architecture. MQTT is a telemetry protocol and uses the publisher/subscriber (pubsub) model, where publisher manages list of resources also known as 'topics' and subscriber can register to 'topics' to obtain information when an event occurs. Similarly, XMPP is extended to implements pubsub model, which implements resources as 'nodes' instead of topics[22]. The SGS architecture provides interfaces to all sink level clients, by supporting these protocols via multi-protocol proxy. Similarly on the other side, the multi-protocol proxy is connected to the gateways as service, which is the Internet facing component of the SGS.

The translation of messages between sink nodes and Internet services is not required when ends, where data is produced and where data is consumed, implement identical messaging mechanism, either REST or pubsub. In cases where the client and server devices have different messaging mechanism, the translation of the message is mandatory at the gateway. The multi-protocol proxy solves the message translation problem via introducing two additional components, message stores and topic router. Each meaningful state of sensor information or resources are described as **topics** and managed by the topic router, which also tracks publisher and subscriber of the topic.

Figure 8 shows message translation between a CoAP client and an MQTT subscriber. When the sensor generates a data or changes its state, the CoAP client sends that change to the SGS as a POST message, which gets captured by the multi-protocol proxy. The proxy aligns that resource with appropriate topic from the topic router and fetches the list of subscriber. The proxy then forwards that message to these subscribers after passing through semantic annotation block.

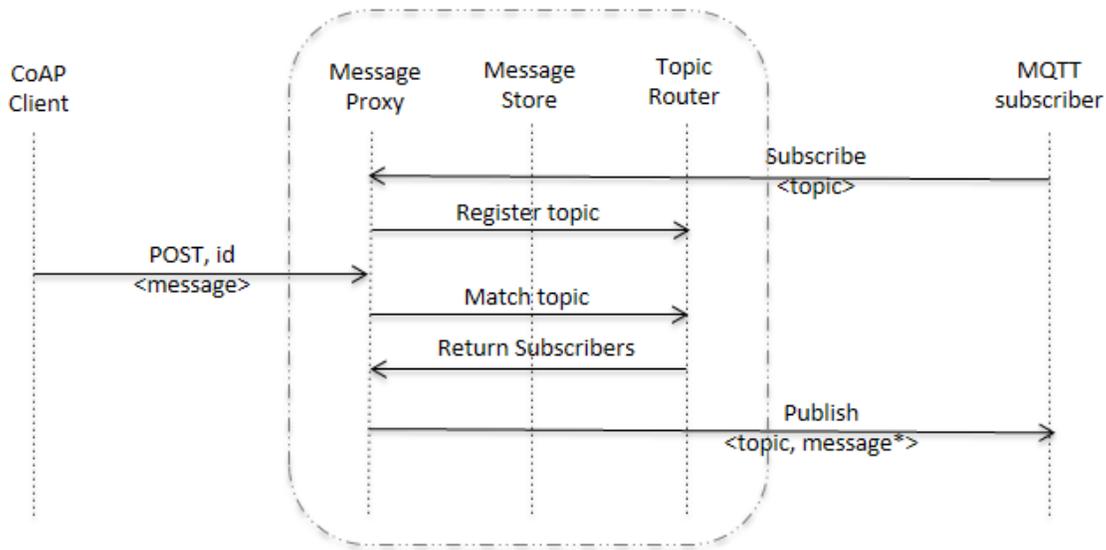

*Figure 8: Message translation from CoAP client to MQTT service.*

Figure 9 shows the translation of messages between a MQTT publisher and REST interface. In this translation process, the message store component is used to buffer the latest message from the publisher to supplement GET request received from the REST interface. The multi-protocol proxy thus solves one of the major interoperability problems at messaging level.

The modular approach of the framework leads to an extensibility, providing interoperability for other IoT protocols such as Advanced Message Queuing Protocol (AMQP) and Data Distribution Service (DDS).

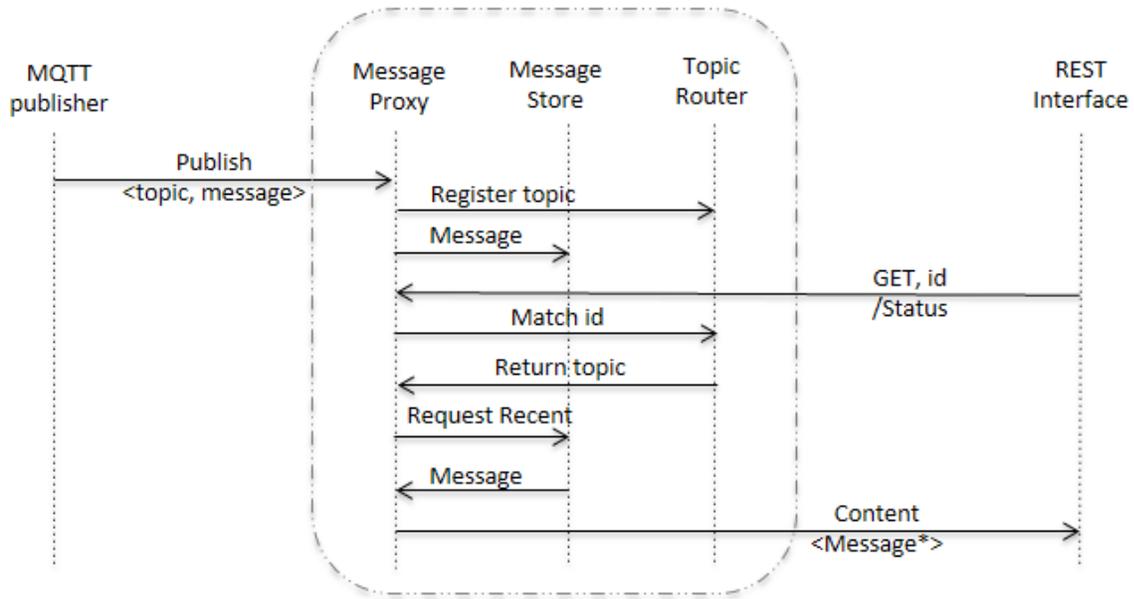

*Figure 9: Message translation from MQTT publisher to REST interface*

## 6. Semantic data annotation

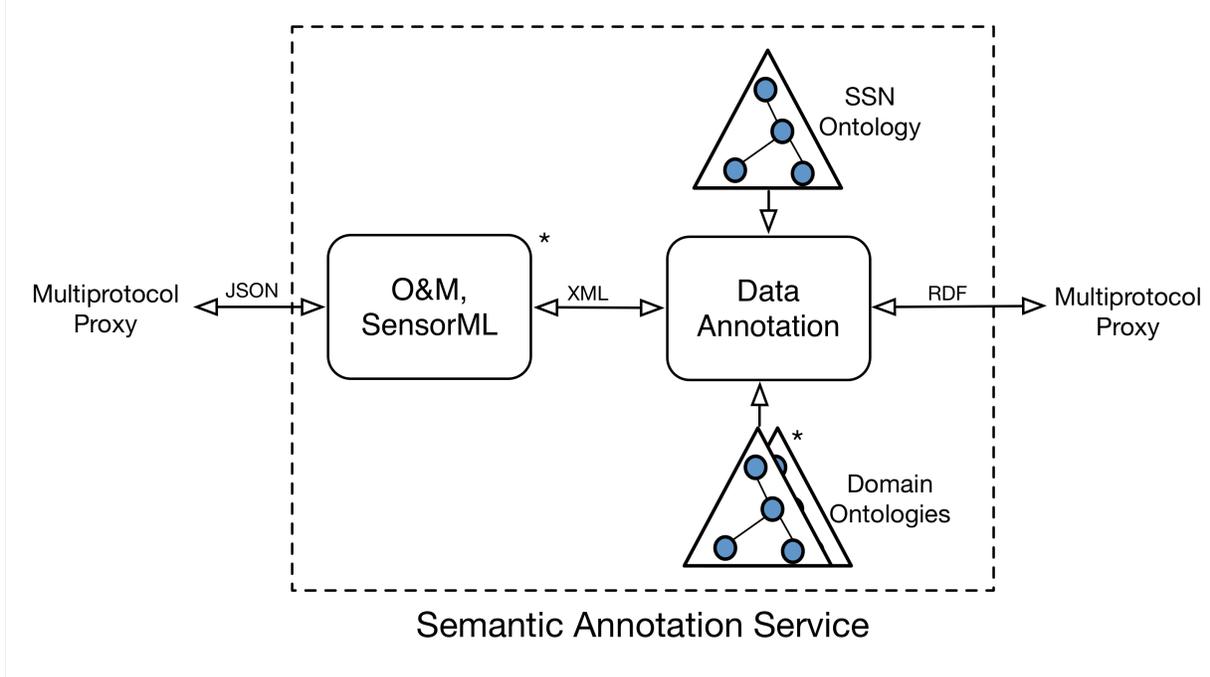

*Figure 10: Semantic data annotation of sensor messages*

The semantic annotation service component process each sensor message received from the sink node before forwarding it further to gateway interface. The annotation process provides standardization at three levels: (1) service description and discovery, (2) sensor and observation description, and (3) domain specific descriptions.

The services based on SOS utilize O&M and SensorML data annotation standards for service description. The SGS annotates the raw sensor data using these OGC standards. This annotation is required for service-oriented systems and for systems to be dynamically discovered by other services. The annotation using OGC standards is optional where number of resources being used are known and well defined.

The semantic sensor and observation description are provided using SSN ontology after annotated with OGC standards. As the primary data model of the proposed architecture, each message is annotated with sensor description using SSN ontology. The semantic sensor description helps other software agents to operate at the level of semantic abstraction, further enabling processing and reasoning over the data[23]. Figure 11 shows a graph describing a temperature sensor observation using various components of SSN ontology.

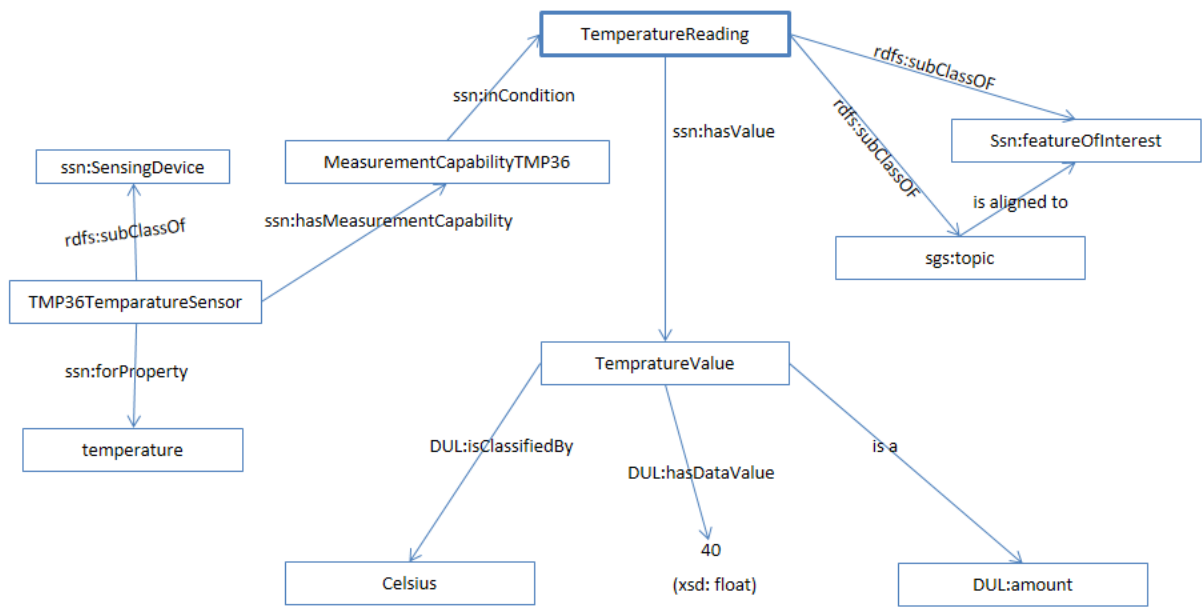

*Figure 11: An example of resource graph for single instance of temperature reading*

For various domain specific applications such as health care, farming, and environmental monitoring, the SGS can be equipped with optional domain specific ontologies. These ontologies describe domain specific concepts to service elements. In a system, which utilizes domain specific ontologies, the SGS is required to communicate those specific ontologies to participating service or other SGSs.

# 7. Gateway service interface

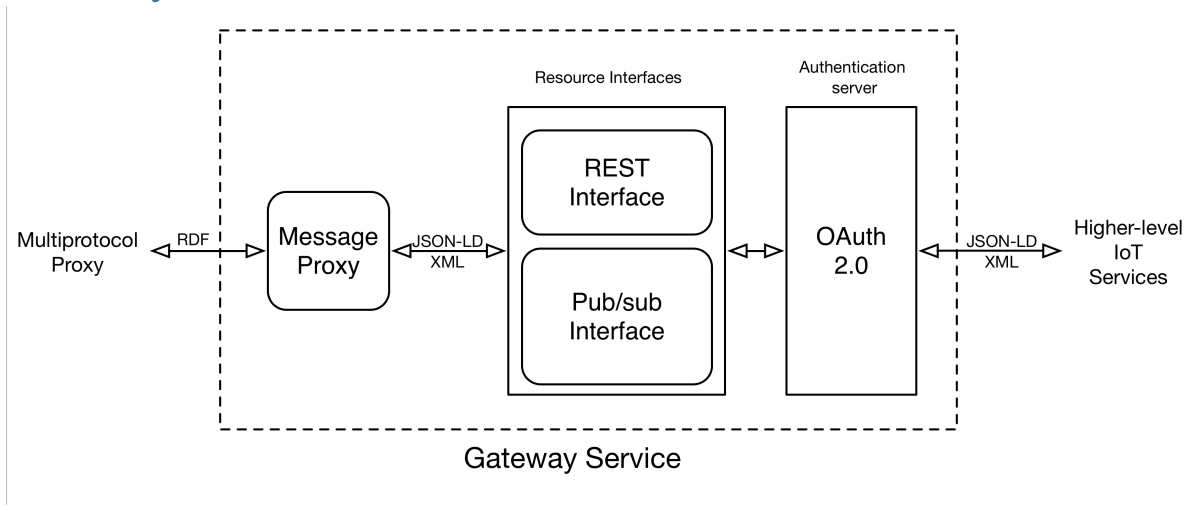

*Figure 12: Gateway as service architecture*

The gateway service is the primary component of the SGS concept as it establishes gateway as the center of the semantic IoT architecture. This component provides service level interoperability for vertical silos of IoT applications keeping physical level implementation independent of cloud based service architecture. The SGS provides endpoint to services using a resource interface via REST and publisher/subscriber mechanism. The MQTT and XMPP protocols are supported via implementing a micro broker in the resource interface. Thus various services can implement response/request and publisher/subscriber mechanism via the SGS component to obtain semantically annotated sensor data. The SGS also provides a layer security via implementing OAuth 2.0 authentication server, which let user decide the private and public resources. Figure 12 shows the gateway service component of the overall SGS architecture, which establishes connectivity between the SGS and higher-level cloud based IoT services.

The cloud based IoT services can be used to provide higher-level knowledge abstractions from the raw sensor data. Various services such as Xively and ThingSpeak provide data analysis and visualization over the collected sensor data but lack implementation of any semantic standards. The semantic annotation of the sensor data obtained from the SGS assists the IoT services to implement analysis and reasoning algorithms. One of the examples of semantic service is the SemSOS implementation, which models sensor and sensor observations utilizing OGC standards[16] with a semantic backend. The SemSOS utilize SSN ontology SSN ontology to model sensors and their observations allowing the implementation of a Semantic reasoner. Figure 13 shows implementation of SemSOS service connected with multiple SGS gateways via Internet. The figure also shows extended version of SemSOS implementation, which includes SSN and domain ontologies to infer sensor description, obtained from SGS implementations. The extended SemSOS can subscribe to the semantic gateways for specific sensor information via selected topics.

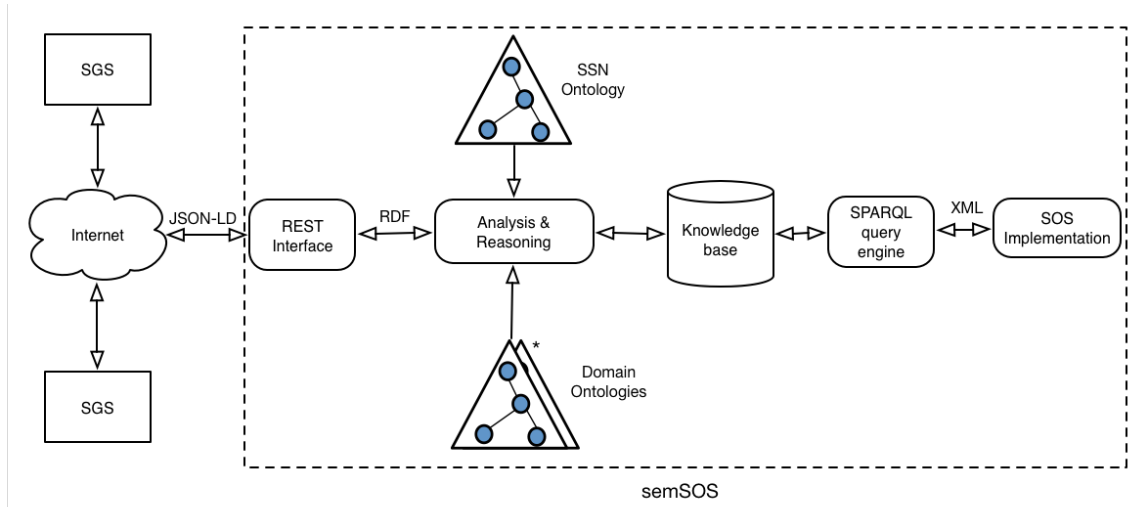

*Figure 13: Higher-level IoT service - SemSOS.*

The semantic annotation using SSN standardizes sensor data making it machine interpretable and thus enabling Machine-to-Machine (M2M) communication. Once data is semantically annotated, various Semantic Web tools can also enable reasoning and higher-level knowledge discovery over sensor data. As the SGS implemented OGC schemas before annotating the sensor data with SSN, the SGS can also provides resource discovery and descriptions/specification for services such as SemSOS. In summary, the sensor data obtained from the multiple SGS is annotated with the standard ontologies enabling service level interoperability.

## 8. Conclusion

Interoperability is one of the major challenges in achieving the vision of Internet of Things. The SGS provides intelligent solution by integrating Semantic Web technologies with existing sensor and services standards. The SGS also provides mechanism to integrate popular IoT application protocol, CoAP and MQTT, to co-exist in a single gateway system. The SGS is integrated with semantic service such as SemSOS to further elevate interoperability at service level. Such a semantic IoT infrastructure can better enable realization of applications spanning the physical world (as observed by IoT), cyberworld (with its rapidly growing data and knowledge about everything in the world, spanning community created Wikipedia to Linked Open Data and repositories of ontologies, as well as its ability to collect and interoperate with all forms of data), and the social world (supporting activities and needs of a person to collective social actions) [24].

Acknowledgements: We acknowledge collaboration with and inputs from the members of EU FP7 project CityPulse http://www.ict-citypulse.eu/ .